\documentclass[useAMS,usenatbib]{mn2e}
\usepackage{graphicx}
\usepackage{amssymb}
\usepackage{amsmath}
\title{Analysing the radio flux density profile of the M31 galaxy: a possible dark matter interpretation}
\author[Chan et al.]{Man Ho Chan$^1$ \thanks{chanmh@eduhk.hk}, Chu Fai Yeung$^1$, Lang Cui$^2$ \thanks{cuilang@xao.ac.cn}, Chun Sing Leung$^3$
\\ $^1$ Department of Science and Environmental Studies, The Education University of Hong Kong, Tai Po, Hong Kong, China \\
$^2$ Xinjiang Astronomical Observatory, Chinese Academy of Sciences, Urumqi, China \\
$^3$ Hong Kong Polytechnic University, Hong Kong, China}

\begin{document}

\date{Accepted XXXX, Received XXXX}

\pagerange{\pageref{firstpage}--\pageref{lastpage}} \pubyear{XXXX}

\maketitle

\label{firstpage}

\date{\today}

\begin{abstract}
Some recent studies have examined the gamma-ray flux profile of our Galaxy to determine the signal of dark matter annihilation. However, the results are controversial and no confirmation is obtained. In this article, we study the radio flux density profile of the M31 galaxy and show that it could manifest a possible signal of dark matter annihilation. By comparing the likelihoods between the archival observed radio flux density profile data and the predicted radio flux density profile contributed by dark matter and stellar emission, we can constrain the relevant dark matter parameters. Specifically, for the thermal annihilation cross section via the $b\bar{b}$ channel, the best-fit value of dark matter mass is $\sim 30$ GeV, which is consistent with the results of many recent studies. We expect that this method would become another useful way to constrain dark matter, which is complementary to the traditional radio analyses and the other indirect detections.
\end{abstract}


\section{Introduction}
In the past several decades, observational data of galaxies and galaxy clusters revealed the existence of dark matter. For example, observations indicate that most galactic rotation curves are flat in the galactic outskirt regions, which contradict to the theoretical predictions based on the luminous mass distributions in galaxies. Also, the dynamical masses of most galaxy clusters probed from their hot gas distributions are much larger than their total luminous masses. Therefore, many astrophysicists believe that a large amount of unknown particles called dark matter exist in the universe. However, none of the particles in the Standard Model can satisfy the unique properties of dark matter - almost no interaction with ordinary matter except gravity.

Many theoretical models suggest that dark matter would self-annihilate to give high-energy gamma rays, electrons, positrons and neutrinos. In particular, some recent analyses of gamma-ray and radio observations have claimed the discoveries of potential signals of annihilating dark matter. For example, using the gamma-ray energy spectrum of our Galaxy or some nearby globular clusters (e.g. 47 Tuc, Omega Centauri) observed from the Fermi-Large Area Telescope, some studies show that dark matter with mass $m \approx 30-40$ GeV annihilating via the $b \bar{b}$ quark channel can best explain the observed spectral shapes \citep{Daylan,Calore,Abazajian,Cholis,Brown,Brown2}. The constrained annihilation cross sections are surprisingly close to the thermal annihilation cross section $\langle \sigma v \rangle \approx 2.2 \times 10^{-26}$ cm$^3$ s$^{-1}$ predicted by standard cosmology \citep{Steigman}. 

On the other hand, indirect detection of dark matter using radio waveband is also good for constraining dark matter parameters. For example, some earlier studies have used the radio data of our Galaxy to constrain dark matter \citep{Blasi,Aloisio,Borriello}. Later studies have started to focus on the Milky Way satellite galaxies such as Large Magellanic Cloud (LMC) \citep{Tasitsiomi,Baltz,Siffert}, nearby galaxies (e.g. M31 and M33 galaxies) \citep{Borriello2,Egorov,Chan} and galaxy clusters \citep{Colafrancesco,Chan2,Chan3,Chan4} to search for dark matter signals. Some recent studies show that certain ranges of dark matter mass can give very good fits to the radio continuum spectra of some galaxy clusters \citep{Chan3,Chan4}.

In particular, most of the previous radio studies have used the total integrated radio emission flux \citep{Blasi,Aloisio}, radio intensity of certain ideal regions \citep{Borriello2,Siffert} or total integrated radio frequency (energy) spectrum (multi-frequency approach) \citep{Tasitsiomi,Borriello,Chan} of a structure to constrain dark matter. Nevertheless, using the radial emission profile can also provide useful constraints for dark matter. For example, some previous studies have shown that the gamma-ray emissions in our Galactic Center trace the dark matter density distribution, which provide an indirect potential evidence of dark matter annihilation \citep{Daylan,Calore}. However, on the contrary, some later studies show that the gamma-ray emissions may probably trace the stellar density profile rather than the dark matter density profile \citep{Bartels}. Therefore, whether the gamma-ray flux profile follows the dark matter distribution in our Galactic Center has now become a controversial issue.

Unfortunately, the resolution of gamma-ray telescopes is not high enough for us to analyse the central gamma-ray flux profiles of other extragalactic targets. Therefore, searching dark matter signals by using the gamma-ray flux profiles of other galaxies is very difficult. Nevertheless, current very large radio telescopes or radio interferometers can obtain radio data with a very high resolution and sensitivity. The resolution can be as small as 10" so that we can observe the radio flux density profiles of galaxies or galaxy clusters. Therefore, radio flux density profile analysis can be used to serve as an independent and complementary study to search for dark matter annihilation signals. In view of this, however, not so much attention has been paid in considering the radial emission profile (the functional form) in radio bands to constrain dark matter. In this article, we use the radio flux density profile of the M31 galaxy as an example to illustrate how this can be done. We show that some potential dark matter annihilation signals can be manifested in the radio flux profile of the M31 galaxy. The best-fit value of $m$ for thermal dark matter generally agrees with the popular range suggested by some gamma-ray studies.

\section{Extracting the radio flux profile data}
We revisit the archival radio data of the central region of the M31 galaxy reported in \citet{Giebubel}. The data were obtained by the combination of the Very Large Array (VLA) and the Effelsberg radio telescope data, which consist of two different frequencies ($\nu=4.85$ GHz and $\nu=8.35$ GHz). The useful data cover sizes of the central region $r \le 1.8$ kpc ($\nu=4.85$ GHz) and $r \le 1.2$ kpc ($\nu=8.35$ GHz). The resolution for both observing frequencies is 15".

We extract the radio flux density per beam size for different positions (RA and DEC) from the radio map published. We take the coordinates RA=00h42m44.3503s and DEC=41$^\circ$16'09" to be the center of the M31 galaxy \citep{Evans}. As the dark matter distribution and the baryonic mass distribution in the bulge region are close to spherically symmetric, we take the azimuthal averaging of the radio flux density (per beam size) in concentric bins for different angular radii $\theta$ from the center. The angular distance for each bin is taken as 26" ($1.27\times 10^{-4}$ rad), which is equivalent to 0.1 kpc at the M31 center (assuming distance to the galaxy $D=785$ kpc \citep{Egorov}). As there are fluctuations in the radio flux density for the same angular radius, we take the $1\sigma$ standard deviation as our radio flux density profile uncertainties. Using this method, we can get the radio flux profiles (radio flux density per beam size against $\theta$) of the central region of the M31 galaxy for two radio frequencies (see Fig.~1). The solid angle of the beam size is $\Delta \Omega =1.66 \times 10^{-8}$ sr.

\section{The dark matter annihilation model}
Dark matter annihilation would give a large amount of high-energy electrons and positrons. These electrons and positrons would emit synchrotron radiation in radio bands when there is a large magnetic field strength. The power for synchrotron emission (with energy $E$) takes the following form \citep{Aloisio,Profumo}:
\begin{equation}
P_{DM}(E,\nu,\vec{r})=\frac{\sqrt{3}e^3}{m_ec^2}B(\vec{r})F(\nu/\nu_c),
\end{equation}
where $\nu_c$ is the critical synchrotron frequency, $B$ is the magnetic field strength and $F$ is the synchrotron kernel function. 

Beside synchrotron radiation cooling, the high-energy electrons and positrons would also cool down via inverse Compton scattering (ICS), Bremsstrahlung radiation and Coulomb loss \citep{Colafrancesco,Egorov}. In particular, synchrotron cooling and the ICS cooling would dominate at the M31 galactic center. Due to the very high cooling rate, most of the high-energy electrons and positrons produced from dark matter annihilation would quickly lose their energy before leaving the central region. However, the actual diffusion length of the high-energy electrons and positrons depends on the diffusion models. For the outer part of a galaxy (e.g. outside the bulge region), the scale of the magnetic field irregularities is comparable to the gyroradius $r_g$ of the electrons and positrons. As a result, the effect of turbulent diffusion is important so that the diffusion of the electrons and positrons is quite efficient in the outer part of a galaxy. Nevertheless, in the central region of a galaxy, the picture is much more uncertain and the scale of magnetic irregularities would be much smaller. Therefore, the diffusion coefficient is significantly reduced \citep{Regis}. Like our Galaxy, the diffusion in the deep central region of the M31 galaxy can be well described by the B\oe hm diffusion model, which means that diffusion effect is not very significant \citep{Regis}. The effective diffusion length of the B\oe hm diffusion model can be estimated by the simple random walk model \citep{Boehm}. The stopping distance of an electron with energy 1 GeV is $d_s \sim \sqrt{r_gct_c} \sim 1$ pc, where $t_c \sim 10^{15}$ s is the cooling time. Therefore, almost all of the electrons and positrons emitted due to dark matter annihilation would be confined in the central region.

By considering the diffusion-cooling equation, as the diffusion term is not important, the equilibrium energy spectrum of the high-energy electrons and positrons become \citep{Borriello,Storm,Egorov}:
\begin{equation}
\frac{dn_e}{dE}= \frac{\langle \sigma v \rangle \rho_{DM}^2}{2m^2b(E)} \int_E^m \frac{dN_e}{dE'}dE',
\end{equation}
where $\rho_{DM}$ is the density profile of dark matter, $b(E)$ is the cooling rate and $dN_e/dE'$ is the injected energy spectrum due to dark matter annihilation, which can be obtained numerically for different annihilation channels \citep{Cirelli}. The radio emissivity contributed by dark matter annihilation is mainly determined by the peak frequency so that the line-of-sight synchrotron radiation flux density (the radio flux density) within a solid angle $\Delta \Omega$ can be simplified by combining Eq.~(1) and Eq.~(2) with the monochromatic approximation \citep{Bertone,Profumo}:
\begin{equation}
S_{DM} \approx \frac{1}{4 \pi \nu} \left[ \frac{9 \sqrt{3} \langle \sigma v \rangle}{2m^2(1+C)} E(\nu)Y(\nu,m) \right] \int \rho_{DM}^2ds \Delta \Omega,
\end{equation}
where $s$ is the line-of-sight distance defined as $r=\sqrt{D^2+s^2-2Ds \cos \theta}$, $Y(\nu,m)=\int_{E(\nu)}^m (dN_e/dE')dE'$ and $E(\nu)=14.6(\nu/{\rm GHz})^{1/2}(B/{\rm \mu G})^{-1/2}$ GeV. Here, $C$ is the correction factor of the ICS cooling contribution. The radiation density due to synchrotron cooling is $\omega_B \approx 9.5$ eV/cm$^3$. For the ICS radiation density, the value for the M31 galaxy central region has not been well-determined. Nevertheless, we know that the value for the Milky Way centre is about $\omega_i=8$ eV/cm$^3$ \citep{Egorov}. Also, the mass of the M31 bulge is about 2.3 times of the mass of the Milky Way bulge \citep{Blana,Karukes}. Therefore, assuming that they have similar luminosity-to-mass ratio, the ICS radiation density for the M31 galaxy central region should be $\omega_i \approx 19$ eV/cm$^3$. Therefore, we have $C \approx 2$ (i.e. ICS cooling is two times larger than the synchrotron cooling). For the central magnetic field, previous observations show that the magnetic field is close to a constant $B=15-19$ $\mu$G for $r=0.2-1$ kpc \citep{Giebubel}. By taking the mean magnetic field strength $B=17$ $\mu$G, the energies at peak frequencies are $E(4.85~\rm GHz)=7.8$ GeV and $E(8.35~\rm GHz)=10.2$ GeV. Besides, standard cosmology predicts that $\langle \sigma v \rangle=2.2 \times 10^{-26}$ cm$^3$/s for $m>10$ GeV if dark matter is thermally produced \citep{Steigman}. We will first follow this standard paradigm to reduce our free parameters in fitting. As a result, $S_{DM}$ would be a function of $\nu$, $m$ and $\theta$ only.

\section{Data analysis}
As the magnetic field strength is almost constant in the region $r=0.2-1.0$ kpc only \citep{Giebubel}, we consider the data from angular radius $\theta=0.0002-0.0012$ rad (i.e. 0.15-0.95 kpc) for our analysis. The dark matter density is assumed to follow the Navarro-Frenk-White (NFW) density profile $\rho_{DM}=\rho_sr_s/r$ (for small $r$) as recent studies show that it can give a very good fit for the M31 galaxy rotation curve data with small uncertainties \citep{Sofue}. The density parameters are $\rho_s=(2.23 \pm 0.24) \times 10^{-3}M_{\odot}$ pc$^{-3}$ and $r_s=(34.6 \pm 2.1)$ kpc \citep{Sofue}. The line-of-sight integral in Eq.~(3) is a function of the angular radius $\theta$. For very small $\theta$, it can be analytically written as (first-order approximation in $\theta$):
\begin{equation}
I(\theta)=\int \rho_{DM}^2ds \approx 2 \sqrt{\pi} \Gamma(3/2) \left(\frac{\rho_s^2r_s^2}{D \theta}\right).
\end{equation}

We fit the data with our dark matter model for various annihilation channels. For each frequency $\nu$, the dark matter mass $m$ would be the only free parameter in the fitting. Generally speaking, the dark matter model can give good fits for the flux profile data for both frequencies. The goodness of fits can be determined by the $\chi^2$ value, which is defined as
\begin{equation}
\chi^2=\sum_i \left[\frac{S_m(\theta_i)-S_o(\theta_i)}{\sigma_o(\theta_i)} \right]^2,
\end{equation}
where $S_m(\theta_i)$, $S_o(\theta_i)$ and $\sigma_o(\theta_i)$ are the data of the modeled flux density, observed flux density and uncertainties of the observed flux density at different angular radii $\theta_i$. Specifically, we find that the $b\bar{b}$ channel with $m \sim 25$ GeV can give the best fit (i.e. smallest $\chi^2$) among all representative popular annihilation channels ($\chi^2=7.1$) (see Table 1 and Fig.~2). We also separate the fit for each radio frequency. The corresponding $\chi^2$ values are 2.6 and 4.5 for 4.85 GHz and 8.35 GHz respectively. 

We also fit the data with the stellar model. We assume that the normal astrophysical emissions such as pulsar emissions are tracing the stellar distribution. The stellar distribution in the bulge is commonly modeled by the Hernquist profile \citep{Sadoun}:
\begin{equation}
\rho_b \propto \frac{1}{r(r+r_b)^3},
\end{equation}
where $r_b=0.61$ kpc for the M31 galaxy. The radio emission due to conventional astrophysical processes (e.g. pulsar emissions) $S_*$ is proportional to the stellar density profile. Therefore, we define the line-of-sight integral for the stellar radio emission as $\tilde{I}(\theta)=\int [r(r+r_b)^3]^{-1}ds$ and we can write $S_*=k\tilde{I}(\theta)$. The proportionality constant $k$ is a free parameter to fit the observed flux profile for each frequency. Since there are two different frequencies in the fits, we have totally two free parameters for fitting. We find that the stellar model can give an overall better fit for the observed flux profiles ($\chi^2=3.2$) compared with the dark matter model (see Table 1), especially for the profile of $\nu=8.35$ GHz ($\chi^2=0.2$). The radio flux density profiles for $S_*$ are shown in Fig.~3. 

Although the stellar model can give an overall better fit compared with the dark matter model, one interesting feature is that the dark matter model gives a better fit for the 4.85 GHz data while the stellar model gives a better fit for the 8.35 GHz. In fact, the radio emission due to dark matter annihilation via the $b\bar{b}$ channel would be more dominant in the low frequency regime because its energy spectrum has a steeper spectral index. Our results might suggest that considering both stellar model and dark matter model would be able to get better fits for both data at 4.85 GHz and 8.35 GHz.

We consider the two-component model $S_{DM}+S_*$ to fit the observed flux profile data. By varying the free parameters $m$ and $k$, we find that a much better overall fit can be obtained ($\chi^2=0.8$) for $m=30$ GeV (see Table 1). The radio flux profiles of the two-component model and the corresponding components are shown in Fig.~4. Compared with the stellar model, the two-component model improves the overall fit with a $1.6 \sigma$ statistical preference. Although it does not indicate a very large signal, including the dark matter annihilation contribution can get an overall better fit. The ratio of dark matter contribution to stellar contribution is approximately ranging from 1:1 to 3:1. Therefore, the central radio flux profile data of the M31 galaxy neither simply trace the stellar profile nor the dark matter density profile, but better trace the stellar plus dark matter density profile.  

We have assumed the thermal annihilation cross section $\langle \sigma v \rangle=2.2 \times 10^{-26}$ cm$^3$ s$^{-1}$ in the above analysis. However, it is possible that dark matter was created non-thermally and the cross section would be different from the standard value. Therefore, we release the cross section as a free parameter so that more general model-independent best-fit values of cross section as a function of dark matter mass could be obtained. In Fig.~5, we plot the best-fit average dark matter annihilation cross section (via $b\bar{b}$ channel) against dark matter mass ($m=20-200$ GeV) based on the two radio flux density profile datasets (the black line), in which the best-fit $\chi^2$ values are same as that in the thermal dark matter scenario ($\chi^2=0.8$). Such a relation can give us the best-fit dark matter mass when we know the values of the annihilation cross section based on different theoretical models or cosmological parameters. Furthermore, if the annihilation cross section is very large, the radio flux density profile contributed by dark matter would probably exceed the flux profile data by a significant amount. Therefore, we can determine the upper limit of the annihilation cross section as a function of dark matter mass within a $2\sigma$ margin. In Fig.~5, we also plot the $2\sigma$ upper limit of the annihilation cross section for $m=20-200$ GeV (the red line).

\begin{figure}
\vskip 10mm
 \includegraphics[width=80mm]{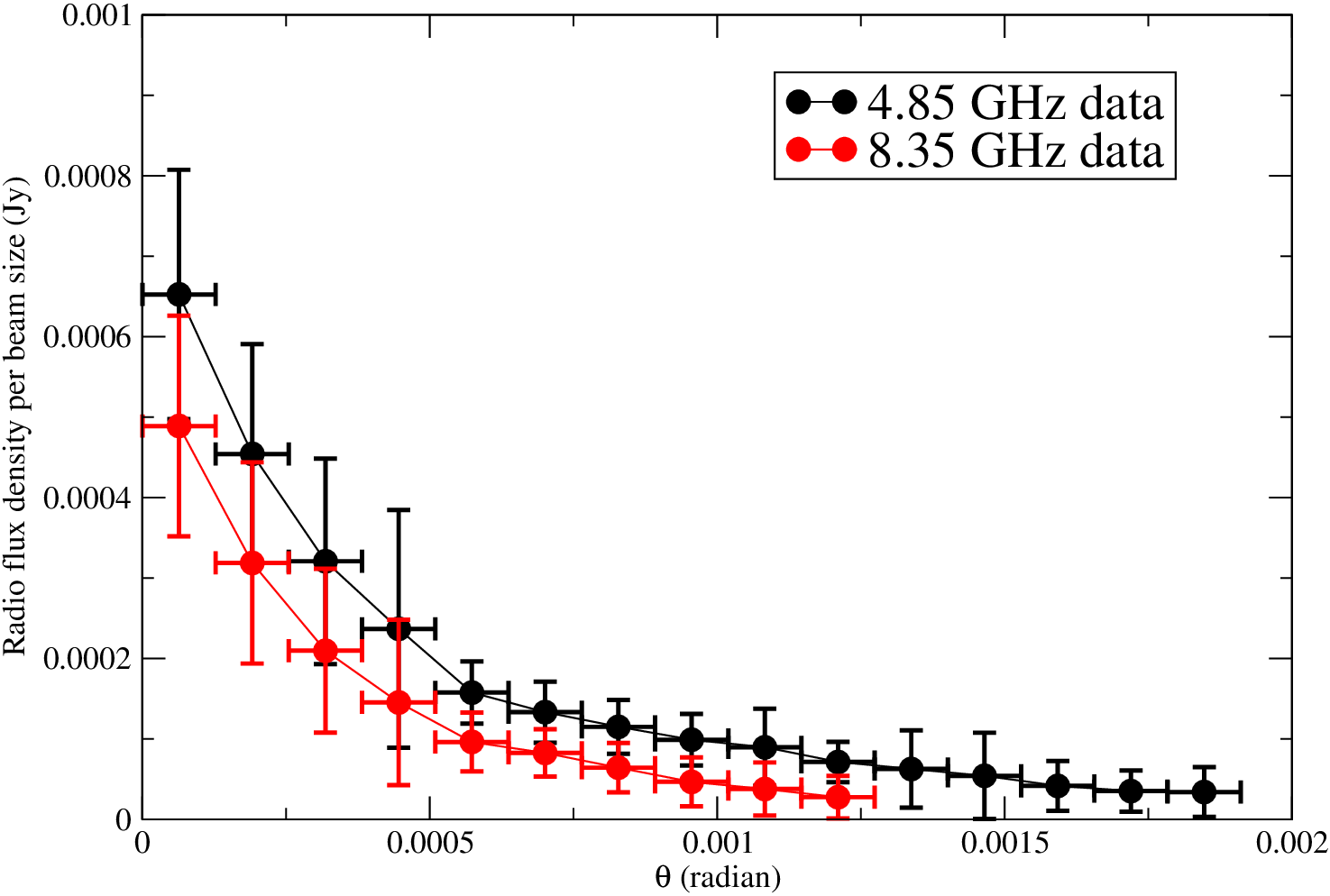}
 \caption{The radio flux density profile data of the central region of the M31 galaxy. The original radio map is taken from \citet{Giebubel}.}
\vskip 10mm
\end{figure}

\begin{figure}
\vskip 10mm
 \includegraphics[width=80mm]{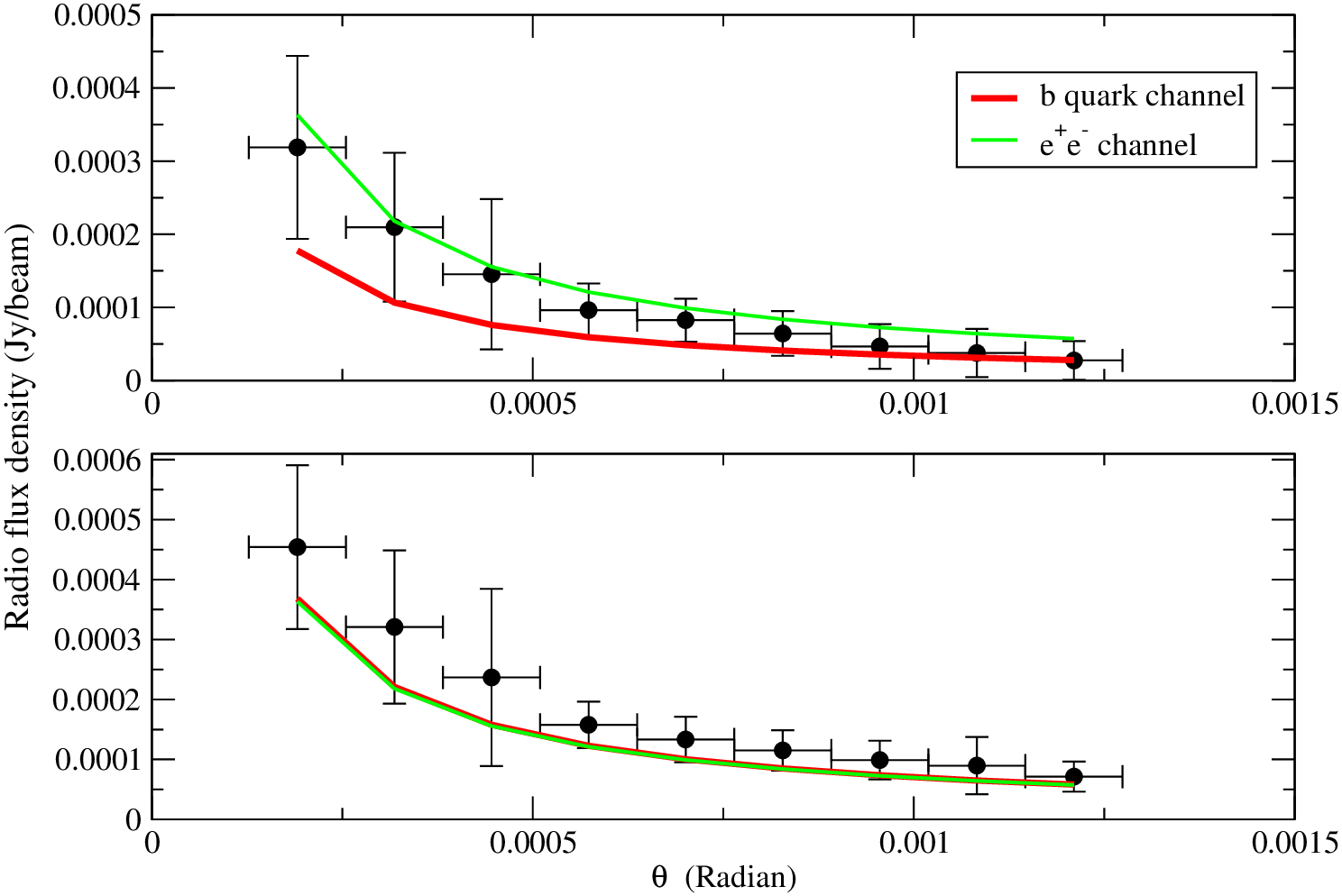}
 \caption{The radio flux density profiles of the dark matter annihilation model (upper: 8.35 GHz; lower: 4.85 GHz). Here $m=25$ GeV for the $b\bar{b}$ channel and $m=185$ GeV for the $e^+e^-$ channel.}
\vskip 10mm
\end{figure}

\begin{figure}
\vskip 10mm
 \includegraphics[width=80mm]{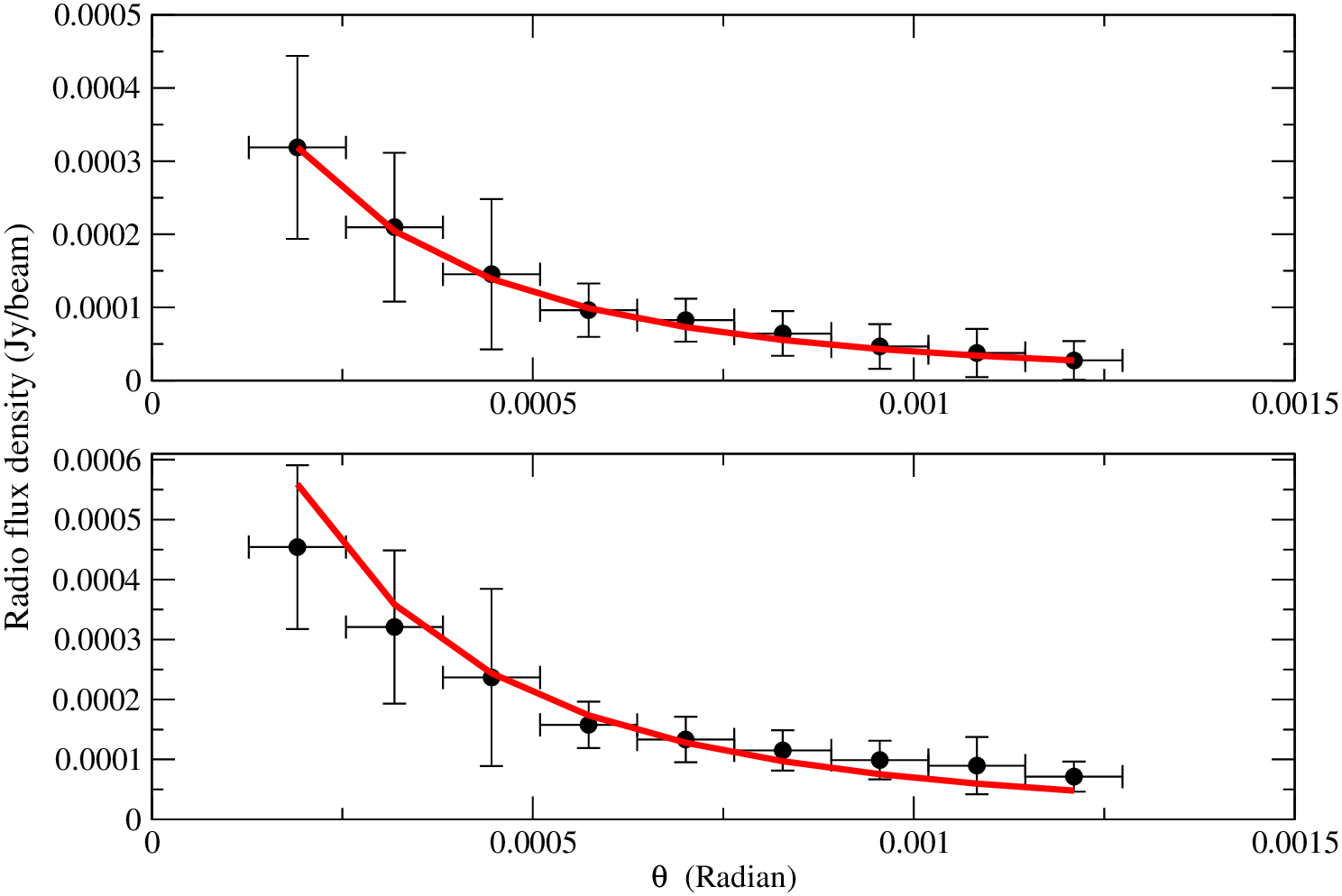}
 \caption{The radio flux density profiles of the stellar model (upper: 8.35 GHz; lower: 4.85 GHz).}
\vskip 10mm
\end{figure}

\begin{figure}
\vskip 10mm
 \includegraphics[width=80mm]{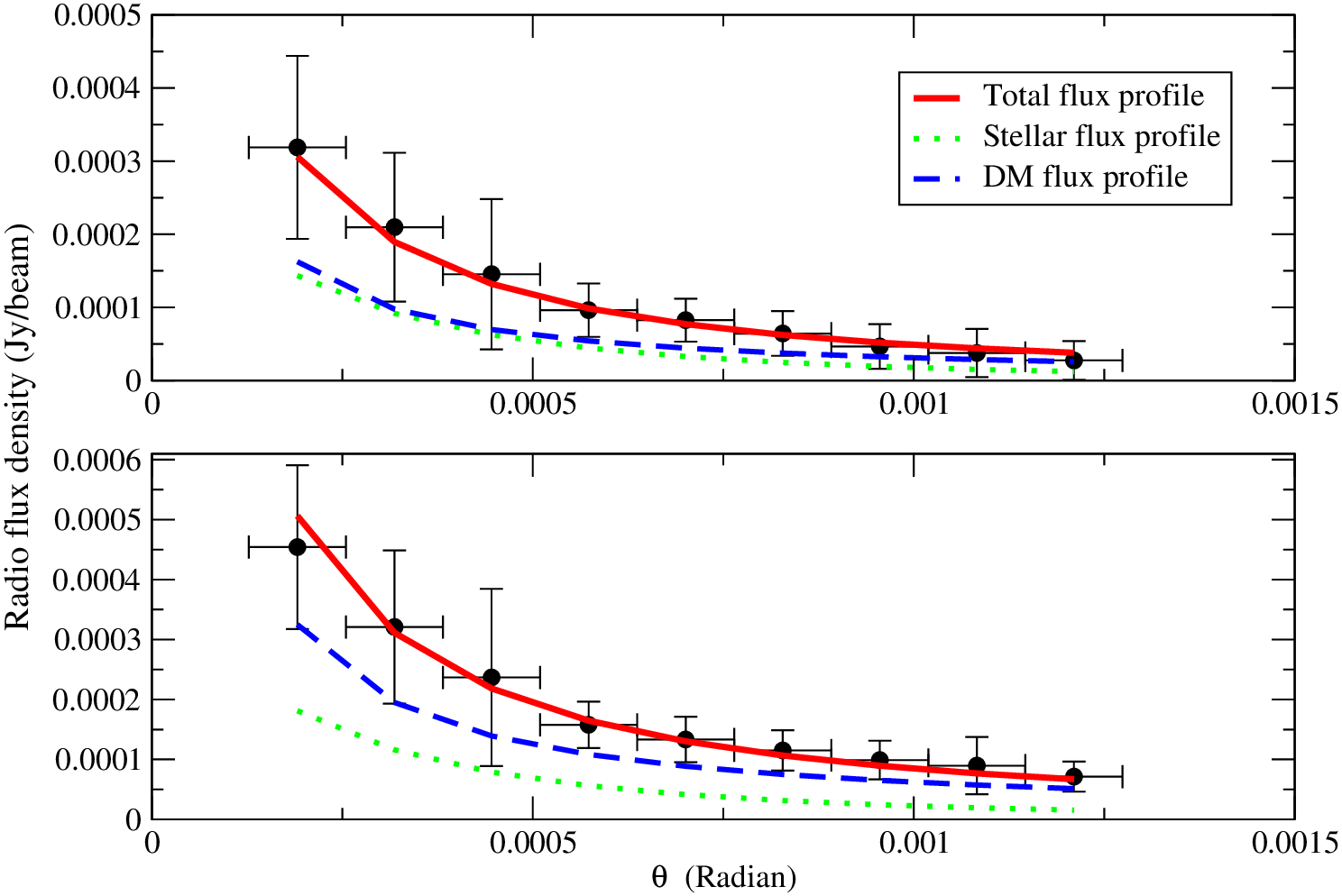}
 \caption{The radio flux density profiles of the two-component model (upper: 8.35 GHz; lower: 4.85 GHz). Here we have assumed $m=30$ GeV annihilating via the $b\bar{b}$ channel.}
\vskip 10mm
\end{figure}

\begin{figure}
\vskip 10mm
 \includegraphics[width=80mm]{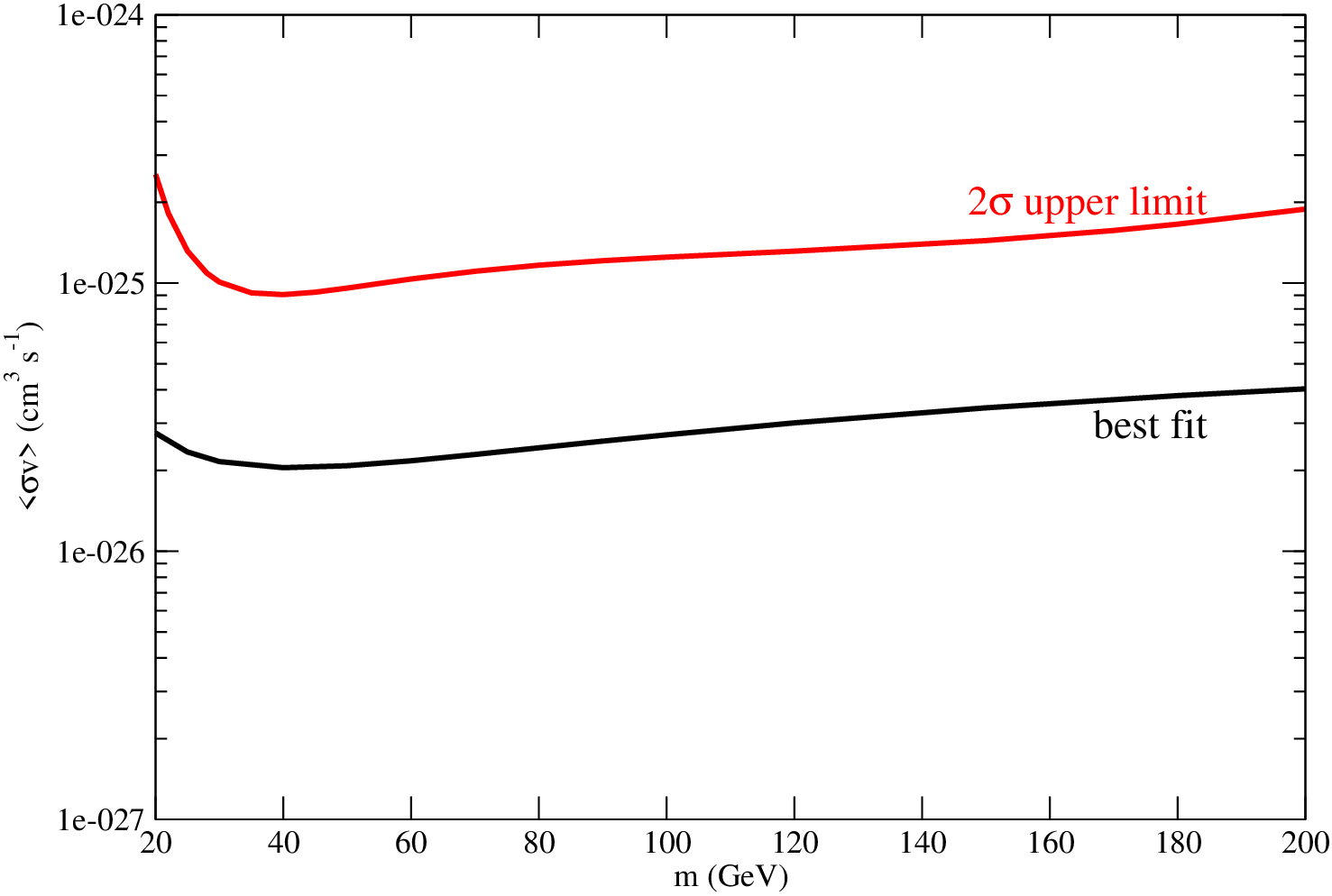}
 \caption{The best-fit (black line) and the 2$\sigma$ upper limit (red line) of the annihilation cross section against dark matter mass. Here we have assumed dark matter annihilating via the $b\bar{b}$ channel.}
\vskip 10mm
\end{figure}

\begin{table}
\caption{The fitting properties for different models.}
 \label{table2}
 \begin{tabular}{@{}lccccc}
  \hline
  Model & Channel &  $m$ & $\chi^2$ & $\chi^2$ & $\chi^2$ \\
        &         & (GeV) & (4.85 GHz) & (8.35 GHz) & (both) \\  
  \hline
  \hline
  $S_*$ & - & - & 2.9 & 0.2 & 3.2 \\
  \hline
  $S_{DM}$ & $b\bar{b}$ & 25 & 2.6 & 4.5 & 7.1 \\
   & $e^+e^-$ & 185 & 5.2 & 4.0 & 9.2 \\
  \hline
  $S_{DM}+S_*$ & $b\bar{b}$ & 30 & 0.5 & 0.3 & 0.8 \\
  \hline
 \end{tabular}
\end{table}

\section{Discussion}
In this article, we discuss an analysis of dark matter annihilation by using the central radio flux density profile of a galaxy. With the observed central radio flux density data of the M31 galaxy at two frequencies, we show that adding dark matter contribution to the background radio emission due to stellar component (the two-component model) can give an overall better fit for the data. The statistical preference of the two-component model is $1.6 \sigma$ compared with the stellar model (the null hypothesis). Therefore, it reveals some potential signal of dark matter annihilation. However, the data and the available frequencies are quite limited so that the resulting flux density profiles have large uncertainties. We expect that a larger set of data with higher resolution and more observing frequencies can provide a much better analysis to confirm the annihilation signal.

In our analysis, with the thermal annihilation cross section, the best-fit $m$ is $\sim 30$ GeV for the two-component model (via $b\bar{b}$ channel). Surprisingly, this value and the annihilation channel is consistent with many recent studies using gamma-ray \citep{Daylan,Brown,Brown2} and radio data \citep{Chan4}. We have assumed that the annihilation cross section follows the thermal annihilation cross section predicted by standard cosmology \citep{Steigman}. Therefore, thermally produced dark matter with mass $\sim 30$ GeV annihilating via $b\bar{b}$ channel has become one of the most probable sets of parameters for dark matter. Further observations and analyses are required to confirm the above claim. Nevertheless, if dark matter was not created thermally, a large range of best-fit dark matter mass is still possible to account for the radio flux density profile of the M31 galaxy.

Previous studies using gamma-ray data of our Galaxy have performed similar analyses. Some studies claim that the gamma-ray flux profile traces dark matter distribution while other studies claim that the gamma-ray flux profile traces stellar distribution. Therefore, it is still a controversial issue. However, the resolution of current gamma-ray detectors is not high enough to get the central gamma-ray flux profiles of other nearby galaxies. Fortunately, the resolution of current radio telescopes is able to fulfill the task. If we can observe the radio flux density profiles of some nearby galaxies with high resolution and different frequencies, some better analyses or clearer signals could be obtained to verify our results. In fact, many previous radio studies mainly focus on constraining dark matter by using the radio intensity of a region (e.g. radio intensity of M33 galaxy or LMC, see \citet{Borriello2,Siffert}) or radio frequency spectrum (multi-frequency approach) (e.g. \citet{Tasitsiomi,Chan}) of galaxies or galaxy clusters. They could identify some ideal regions with the lowest signal to noise ratio for dark matter detection in the radio wavebands. Some other studies have obtained the radio sky map of our Galaxy and used the sky map data to constrain dark matter \citep{Borriello}. However, these studies have not explicitly examined the likelihoods between the observed radial emission profile and the predicted radial emission profile contributed by dark matter annihilation. In our study, we show that using high resolution radio density flux profile at the central region of a galaxy is good for constraining dark matter. Therefore, observing and analysing the radio flux density profile (i.e. the radial emission profile) would be another important way to detect the signal of dark matter annihilation and constrain dark matter properties, which are complementary to the traditional radio analyses (using total integrated radio flux or frequency spectrum) \citep{Blasi,Aloisio,Tasitsiomi,Borriello,Chan}, cosmic-ray (including gamma-ray) analyses \citep{Ackermann,Abdallah,Ambrosi,Aguilar} and neutrino analyses \citep{Albert} of dark matter annihilation.

\section{Acknowledgements}
The work described in this paper was supported by a grant from the Research Grants Council of the Hong Kong Special Administrative Region, China (Project No. EdUHK 28300518) and the Internal Research Fund from The Education University of Hong Kong (RG 2/2019-2020R). Lang Cui thanks for the support from the National Key R\&D Program of China (No. 2018YFA0404602), the National Natural Science Foundation of China (NSFC grant No. 61931002 \& U1731103) and the Youth Innovation Promotion Association of the CAS (No. 2017084).

\section{Data availability statement}
The data underlying this article will be shared on reasonable request to the corresponding author.

\label{lastpage}


\begin{thebibliography}{}
\bibitem[Abazajian \& Keeley (2016)]{Abazajian} Abazajian K. N. \& Keeley R. E., 2016, Phys. Rev. D 93, 083514.
\bibitem[Abdallah et al. (2016)]{Abdallah} Abdallah H. {\it et al.} [H.E.S.S. Collaboration], 2016, Phys. Rev. Lett. 117, 111301.
\bibitem[Ackermann et al. (2015)]{Ackermann} Ackermann M. {\it et al.}, 2015, Phys. Rev. Lett. 115, 231301.
\bibitem[Aguilar et al. (2019)]{Aguilar} Aguilar M. {\it et al.}, 2019, Phys. Rev. Lett. 122, 041102.
\bibitem[Albert et al. (2020)]{Albert} Albert A. {\it et al.}, 2020, arXiv:2003.06614.
\bibitem[Aloisio, Blasi \& Olinto (2004)]{Aloisio} Aloisio R., Blasi P. \& Olinto A. V., 2004, J. Cosmol. Astropart. Phys. 05, 007.
\bibitem[Ambrosi et al. (2017)]{Ambrosi} Ambrosi G. {\it et al.}, 2017, Nature 552, 63.
\bibitem[Baltz \& Wai (2004)]{Baltz} Baltz E. A. \& Wai L., 2004, Phys. Rev. D 70, 023512.
\bibitem[Bartels et al. (2018)]{Bartels} Bartels R., Storm E., Weniger C. \& Calore F., 2018, Nature Astron. 2, 819.
\bibitem[Bertone et al. (2009)]{Bertone} Bertone G., Cirelli M., Strumia A. \& Taoso M., 2009, J. Cosmol. Astropart. Phys. 03, 009.
\bibitem[Bla\~na et al. (2018)]{Blana} Bla\~na Diaz M. {\it et al.}, 2018, Mon. Not. R. Astron. Soc. 481, 3210.
\bibitem[Blasi, Olinto \& Tyler (2003)]{Blasi} Blasi P., Olinto A. V. \& Tyler C., 2003, Astropart. Phys. 18, 649.
\bibitem[B\oe hm et al. (2004)]{Boehm} B\oe hm C., Hooper D., Silk J., Casse M. \& Paul J., 2004, Phys. Rev. Lett. 92, 101301.
\bibitem[Borriello, Cuoco \& Miele (2009)]{Borriello} Borriello E., Cuoco A. \& Miele G., 2009, Phys. Rev. D 79, 023518.
\bibitem[Borriello et al. (2010)]{Borriello2} Borriello E., Longo G., Miele G., Siffert B. B., Tabatabaei F. S. \& Beck R., 2010, Astrophys. J. 709, L32.
\bibitem[Brown et al. (2018)]{Brown} Brown A. M., Lacroix T., Lloyd S., B\oe hm C. \& Chadwick P., 2018, Phys. Rev. D 98, 041301.
\bibitem[Brown et al. (2019)]{Brown2} Brown A. M., Massey R., Lacroix T., Strigari L. E., Fattahi A. \& B\oe hm C., arXiv:1907.08564.
\bibitem[Calore et al. (2015)]{Calore} Calore F., Cholis I., McCabe C. \& Weniger C., 2015, Phys. Rev. D 91, 063003.
\bibitem[Chan (2018)]{Chan} Chan M. H., 2018, Mon. Not. R. Astron. Soc. 474, 2576.
\bibitem[Chan \& Lee (2019)]{Chan3} Chan M. H. \& Lee C. M., 2019, Phys. Dark Univ. 26, 100355.
\bibitem[Chan \& Lee (2021)]{Chan4} Chan M. H. \& Lee C. M., 2021, Mon. Not. R. Astron. Soc. 500, 5583 (arXiv:1912.03640).
\bibitem[Chan et al. (2020b)]{Chan2} Chan M. H., Lee C. M., Ng C.-Y. \& Leung C. S., 2020b, Astrophys. J., 900, 126 (arXiv:2007.06547).
\bibitem[Cholis, Linden \& Hooper (2019)]{Cholis} Cholis I., Linden T. \& D. Hooper, 2019, Phys. Rev. D 99, 103026.
\bibitem[Cirelli et al. (2011)]{Cirelli} Cirelli M. et al., 2011, J. Cosmol. Astropart. Phys. 03, 051.
\bibitem[Colafrancesco, Profumo \& Ullio (2006)]{Colafrancesco} Colafrancesco S., Profumo S. \& Ullio P., 2006, Astron. Astrophys. 455, 21.
\bibitem[Colafrancesco, Marchegiani \& Beck (2015)]{Colafrancesco2} Colafrancesco S., Marchegiani P. \& Beck G., 2015, J. Cosmol. Astrophys. Phys. 02, 032.
\bibitem[Daylan et al. (2016)]{Daylan} Daylan T., Finkbeiner D. P., Hooper D., Linden T., Portillo S. K. N., Rodd N. L. \& Slatyer T. R., 2016, Phys. Dark Univ. 12, 1.
\bibitem[Egorov \& Pierpaoli (2013)]{Egorov} Egorov A. E. \& Pierpaolo E., 2013, Phys. Rev. D 88, 023504.
\bibitem[Evans et al. (2010)]{Evans} Evans I. N. {\it et al.}, 2010, Astrophys. J. Supp. 189, 37.
\bibitem[Gie\ss \"ubel \& Beck (2014)]{Giebubel} Gie\ss \"ubel R. \& Beck R., 2014, Astron. Astrophys. 571, A61.
\bibitem[Karukes et al. (2020)]{Karukes} Karukes E. V., Benito M., Iocco F., Trotta R. \& Geringer-Sameth A., 2020, J. Cosmol. Astropart. Phys. 05, 033.
\bibitem[Profumo \& Ullio (2010)]{Profumo} Profumo S. \& Ullio P., {\it Particle Dark Matter: Observations, Models and Searches}, ed. G. Bertone, Cambridge: Cambridge University Press, chapter 27 (2010).
\bibitem[Regis \& Ullio (2008)]{Regis} Regis M. \& Ullio P., 2008, Phys. Rev. D 78, 043505.
\bibitem[Sadoun, Mohayaee \& Colin (2014)]{Sadoun} Sadoun R., Mohayaee R. \& Colin J., 2014, Mon. Not. R. Astro. Soc. 442, 160.
\bibitem[Siffert et al. (2011)]{Siffert} Siffert B. B., Limone A., Borriello E., Longo G. \& Miele G., 2011, Mon. Not. R. Astron. Soc. 410, 2463.
\bibitem[Sofue (2015)]{Sofue} Sofue Y., 2015, Publ. Astron. Soc. Jpn. 67, 759.
\bibitem[Steigman, Dasgupta \& Beacom (2012)]{Steigman} Steigman G., Dasgupta B. \& Beacom J. F., 2012, Phys. Rev. D 86, 023506.
\bibitem[Storm et al. (2013)]{Storm} Storm E., Jeltema T. E., Profumo S. \& Rudnick L., 2013, Astrophys. J. 768, 106.
\bibitem[Tasitsiomi, Siegal-Gaskins \& Olinto (2004)]{Tasitsiomi} Tasitsiomi A., Siegal-Gaskins J. M. \& Olinto A. V., 2004, Astropart. Phys. 21, 637.
\end{thebibliography}
\end{document}